\documentclass[aps,prl,reprint,superscriptaddress]{revtex4-1}

\usepackage{graphicx}
\usepackage{caption}
\usepackage{subcaption}
\usepackage{lmodern}
\usepackage{amsmath}
\usepackage{amsfonts}
\usepackage{xfrac}
\usepackage{array}

\begin{document}


\title{Improving predictability of time series using maximum entropy methods}



\author{Gregor Chliamovitch}
\email{Gregor.Chliamovitch@unige.ch}
\affiliation{Department of Theoretical Physics, University of Geneva, Switzerland}
\affiliation{Department of Computer Science, University of Geneva, Switzerland}

\author{Bastien Chopard}
\affiliation{Department of Computer Science, University of Geneva, Switzerland}

\author{Alexandre Dupuis}
\affiliation{Department of Computer Science, University of Geneva, Switzerland}

\author{Anton Golub}
\affiliation{Department of Computer Science, University of Geneva, Switzerland}


\date{\today}

\begin{abstract}
We discuss how maximum entropy methods may be applied to the reconstruction of Markov processes underlying empirical time series and compare this approach to usual frequency sampling. It is shown that, at least in low dimension, there exists a subset of the space of stochastic matrices for which the MaxEnt method is more efficient than sampling, in the sense that shorter historical samples have to be considered to reach the same accuracy. Considering short samples is of particular interest when modelling smoothly non-stationary processes, for then it provides, under some conditions, a powerful forecasting tool. The method is illustrated for a discretized empirical series of exchange rates.
\end{abstract}

\pacs{}
\keywords{}

\maketitle


\section{}

A prominent issue of modern science is that experimental data are collected at ever higher frequencies, which increasingly challenges our ability to shape sensible pictures from such data flows and makes necessary to develop statistical tools providing a \emph{live} comprehension of phenomena. We intend here to adress a simple instance of this question: assuming we observe a time series driven by some unknown Markov dynamics, what is the most efficient way to rebuild this underlying dynamics ? In a second time, we shall re-consider the question when the time series is not \emph{generated} by a Markov process but only \emph{modelled} by it, and extend it to the non-stationary case.

The naive way to tackle this problem is to consider a historical window from which the transition probabilities are deduced by counting the transitions which occurred. We shall instead resort to a class of statistical methods known as maximum entropy (MaxEnt) methods. This approach was introduced to physicists by Jaynes \cite{Jaynes1957a, Jaynes1957b, Cover2006} as an attempt to free Statistical Mechanics from physical assumptions going beyond first principles (thereby we mean any assumption related to the ergodic hypothesis) and to disentangle its physical and statistical aspects. The purpose of the method is to build generic probability distributions on the basis of partial knowledge, using the criterion that the appropriate distribution is the one which has the largest entropy while satisfying a set of constraints. A classical example is Gibbs' canonical distribution, which is the maximum entropy distribution corresponding to constrained mean energy. Much more recently, this method has been applied successfully to other kinds of systems, from neuroscience to linguistics \cite{Schneidman2003, Schneidman2006, Stephens2010, Stephens2013}.

Although this seems to be scarcely done \cite{VDS2009}, this method may be transferred to the realm of Markov stochastic processes. Adaptations are however necessary, since considering the joint entropy of some given subsequence of a process would appear arbitrary; one should instead maximize the entropy rate (or entropy per sign) $h= - \sum_{ij} \mu_i W_{ij} \ln W_{ij}$, where $W_{ij}$ denotes the probability of transitioning from state $i$ to state $j$, and $\mu$ the stationary distribution. Unfortunately, $\mu$ itself depends on $\mathbf{W}$ which makes the calculations considerably more involved. A way out \cite{VDS2009} is to maximize instead

\begin{equation}
\eta := - \sum_{ij} p_i W_{ij} \ln W_{ij}
\end{equation}
where $\mathbf{p}$ is some distribution which is kept independent. One ensures that $\mathbf{p}$ eventually \emph{is} the stationary  distribution by imposing the detailed balance condition $p_i W_{ij} = p_j W_{ji}$ $\forall i,j$. In other words, we end up with the following three structural constraints:

\begin{equation}
\sum_i p_i = 1 \quad , \quad \sum_{j} W_{ij} = 1 \quad , \quad p_i W_{ij}=p_j W_{ji} .
\label{trivialconstraints}
\end{equation}
A simple quantity one is likely to measure in order to get a non-trivial process is the autocorrelation of the time series, and it happens that considering \emph{one-step} autocorrelation $A$ makes calculations especially tractable. We therefore constrain the one-step autocorrelation, namely \footnote{Defining autocorrelation without centering nor normalization is usual in the field of signal processing.}

\begin{equation}
\sum_{ij} x_i x_j p(i,t;j,t+1) = \sum_{ij} x_i x_j p_i W_{ij} = A .
\label{constraintAC}
\end{equation}

For constraints given by \eqref{trivialconstraints} and \eqref{constraintAC}, the MaxEnt method requires maximizing

\begin{align}
\eta + \alpha \sum_i p_i + \sum_i \beta_i \sum_{j} W_{ij} + \sum_{ij} \gamma_{ij} (p_i W_{ij} - p_j W_{ji}) \notag \\
+ \lambda \sum_{ij} x_i x_j p_i W_{ij} ,
\label{VariationalProblem}
\end{align}
where $\alpha$, $\beta_i$, $\gamma_{ij}$ and $\lambda$ are the multipliers associated with the respective constraints.

Deriving \eqref{VariationalProblem} with respect to $p_i$ and $W_{ij}$ and equating to zero results in the system

\begin{equation}
\frac{W_{ii}^{ME}}{W_{jj}^{ME}} = \exp \left( \lambda i^2 - \lambda j^2 \right)
\label{system1}
\end{equation}

\begin{equation}
\frac{W_{ii}^{ME} W_{jj}^{ME}}{W_{ij}^{ME} W_{ji}^{ME}} = \exp \left( \lambda (i-j)^2 \right) .
\label{system2}
\end{equation}
In general the $W_{ij}^{ME}$'s have to be found numerically which may be computationally difficult. In order to focus on the aspects of the method which are related to statistical inference, we shall restrict ourselves to 2- and 3-state cases, for which $\mathbf{W}^{ME}$ is easily and quickly found. These cases are enough to cover many situations of interest as many models deal with such binarized or ternarized states (for financial models like the one considered below, this would correspond to \emph{up}, \emph{down} or \emph{flat} market moves \cite{Voit2005}).

We now focus on the case of two states encoded as $\pm1$. Letting $A$ denote the autocorrelation of the process, the MaxEnt transition matrix is

\begin{equation}
\mathbf{W}^{(ME)} = \left(\begin{array}{cc}
\frac{1+A}{2} &\frac{1-A}{2} \\
\frac{1-A}{2} & \frac{1+A}{2} \\
\end{array}
\right) .
\label{2-states matrix}
\end{equation}
We now prove that there exists a subset of the space of $2 \times 2$ stochastic matrices for which the MaxEnt method is more efficient in estimating $\mathbf{W}$ when we only have short samples at our disposal. We detail the calculations for the coefficient $W_{--}$, the other three being similar.

 Since the sample autocorrelation of a well-behaved process obeys a central limit theorem \cite{Brockwell1991}, we make the assumption that the sample autocorrelation $A^{(n)}$ measured from a sample of size $n$ is distributed normally according to $\mathcal{N}(A, n^{-1})$. According to \eqref{2-states matrix}, it follows that the error commited on the estimation of $W_{--}$ using the MaxEnt method is distributed as $\mathcal{N}(\frac{1+A}{2}-W_{--}, (4n)^{-1})$. The absolute value of this error thus obeys a folded normal distribution, which has mean and standard deviation given by \cite{Leone1961}:

\begin{equation}
\langle \vert \Delta_{ME} W_{--} \vert \rangle^{(n)} = \frac{e^{- 2 n \mu_{--}^2}}{\sqrt{2\pi n}} + \mu_{--} \left( 1 - 2\Phi \left( - 2 \sqrt{n} \mu_{--} \right) \right)
\end{equation}

\begin{equation}
\sigma^{(n)} (\vert \Delta_{ME} W_{--} \vert) = \sqrt{ \mu_{--}^2 + \frac{1}{4n} - \left( \langle \vert \Delta W_{--} \vert \rangle_{ME}^{(n)} \right)^2 },
\end{equation}
where $\mu_{--}=\frac{1+A}{2}-W_{--}$ and $\Phi$ denotes the standard normal cumulative distribution.

We can similarly provide an estimation of the error commited when estimating $W_{--}$ by frequency sampling. It can be shown that the coefficient sampled from a window of size $n$ is distributed normally according to $\mathcal{N}(W_{--}, \frac{W_{--} (1-W_{--})}{n p_{-}})$ where $p_{-}$ denotes the stationary probability of being in state $-1$, which in the current setting is given by $p_{-}= \frac{1-W_{++}}{2-W_{--}-W_{++}}$. Following the same steps as previously, the sampled absolute error on $W_{--}$ has mean and deviation:

\begin{equation}
\langle \vert \Delta_S W_{--} \vert \rangle^{(n)} = \sqrt{\frac{2 W_{--} (1-W_{--})}{\pi n p_{-}}}
\end{equation}

\begin{equation}
\sigma^{(n)} (\vert \Delta_S W_{--} \vert) = \sqrt{ \left( 1- \frac{2}{\pi} \right) \frac{W_{--}(1-W_{--})}{n p_{-}} }.
\end{equation}

We are therefore led to define the \emph{accuracy gain}

\begin{equation}
\Delta_{--}^{(n)} := \langle \vert \Delta_S W_{--} \vert) \rangle^{(n)} - \langle \vert \Delta_{ME} W_{--} \vert \rangle^{(n)}
\end{equation}
which is positive when the MaxEnt method provides a better estimation of $W_{--}$ than frequency sampling does for samples of size $n$. Let $n_c^{--}$ be the value of $n$ above which $\Delta_{--}^{(n)}$ becomes negative. Though $n_c^{ij}$ depends on the coefficient, one may wish to define a global $n_c$ for the $\mathbf{W}$ matrix considered. While a conservative option is to choose the minimum over all coefficients, we shall rather tolerate a poor estimation of one of the coefficients as long as the corresponding transitions occur scarcely and therefore define $n_c(\mathbf{W})$ as the sum of all $n_c^{ij}$'s, weighted by the stationary distribution.

\begin{figure}
\includegraphics[scale=0.45]{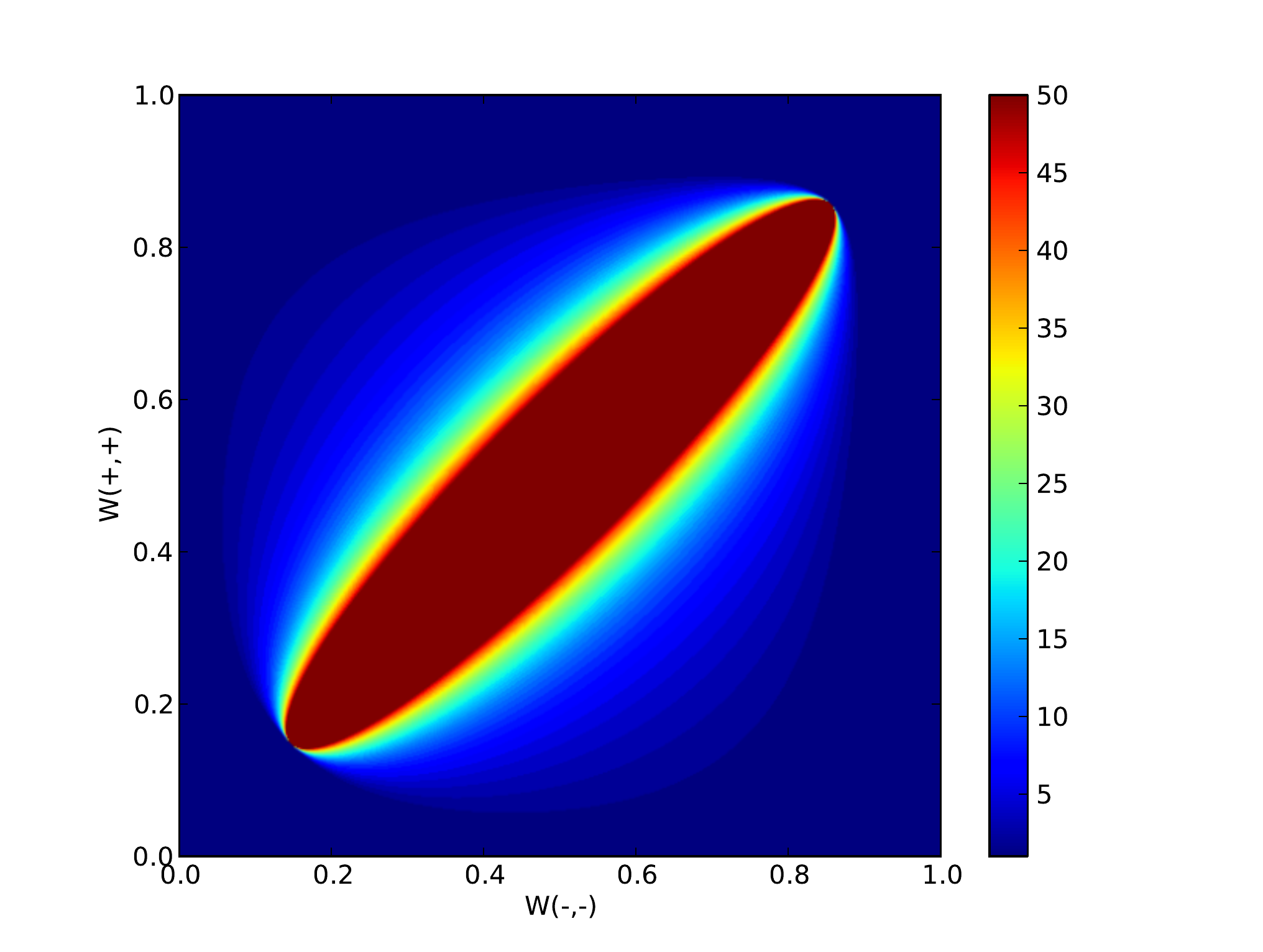}
\caption{\small $n_c(\mathbf{W})$ plotted over the space of 2-states stochastic matrices parametrized by $W_{--}$, $W_{++}$}
\label{Fig1}
\end{figure}

The quantity $n_c(\mathbf{W})$ is found numerically and plotted in figure \ref{Fig1} over the space of $2\times2$ stochastic matrices parametrized by $(W_{--}, W_{++})$. Note that $n_c$ is large close to the diagonal but decays when one gets remote from the diagonal. This means that a matrix which is ``compatible'' with structure \eqref{2-states matrix} is better estimated using MaxEnt, while for matrices which do not fit well into this structure the sampling will turn out to be more adequate.

Denoting $M(n)$ the set of matrices such that $n_c(\mathbf{W}) \geq n$ and $\mu(n)$ the relative size of $M(n)$ compared to $M(0)$ (the space of all $2 \times 2$ stochastic matrices), then the relevance of the MaxEnt approach for a given state space obviously will depend critically on the function $\mu(n)$. In the 2-state case, one can read from figure \ref{Fig1} that $M(50)$ is concentrated in a neighbourhood around the diagonal so that $\mu(50) \approx 0.15$, which means that for samples of size smaller than $n=50$ the MaxEnt estimate is better than the frequency sampling estimate for $15\%$ of all possible processes. One should however note that processes on which one might want to apply the method are unlikely to be scattered randomly over $[0,1]^2$, but will rather be processes having a large entropy, that is low predictability. This tends to focus our interest on the central area of $[0,1]^2$ and increase the effective $\mu(n)$.

Though obviously $\mu(n)$ changes with the dimensionality of the state space, it is difficult to set up a general argument showing how $\mu(n)$ evolves when the state space gets larger. Intuitively, an efficient frequency sampling in a high-dimensional space should require very long samples, so that one would expect the MaxEnt to outperform sampling in high dimensions.

\begin{figure}
\includegraphics[scale=0.5]{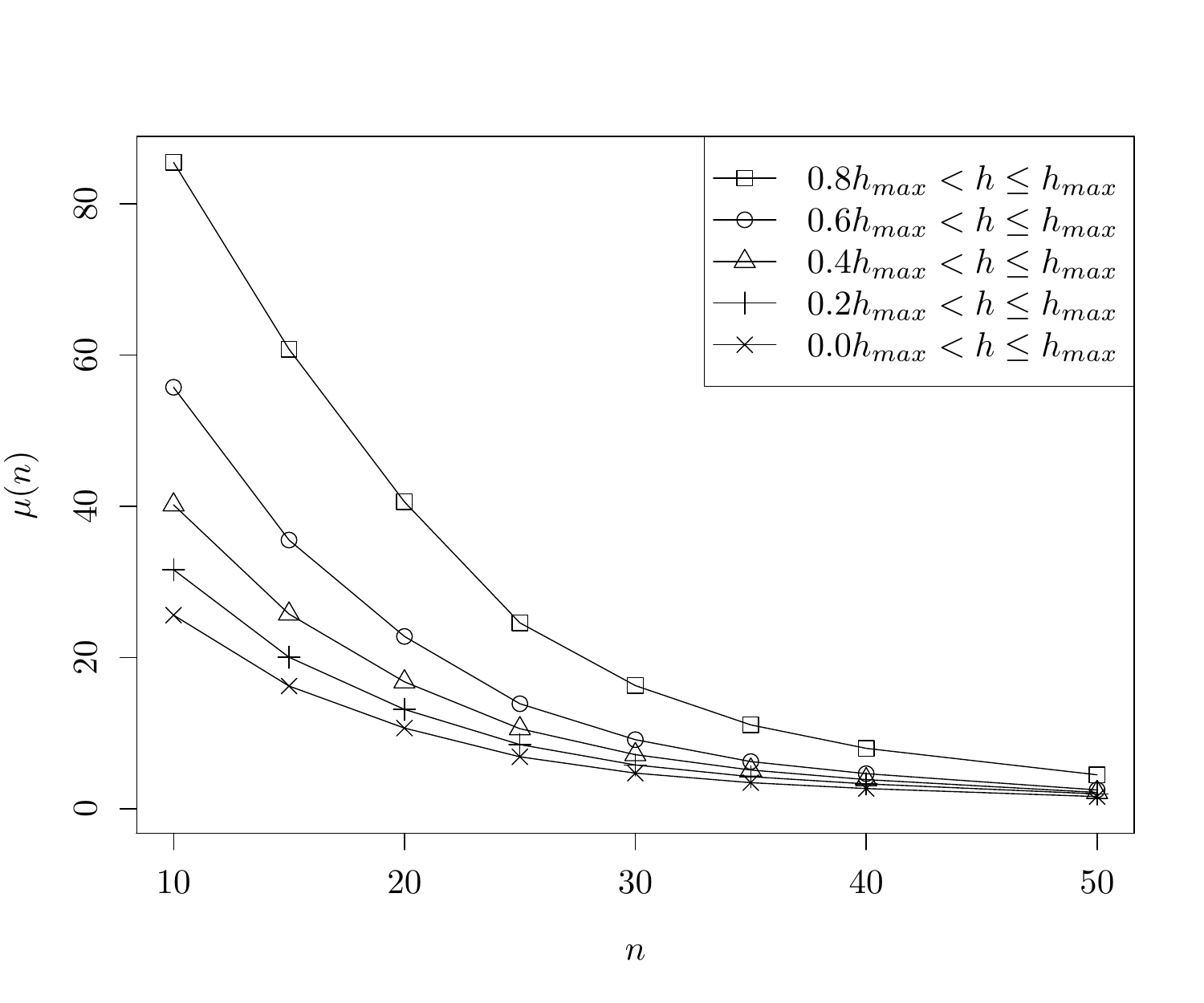}
\caption{\small Percentage of favourable 3-state matrices vs. sample size. Curves are plotted for each cumulated quintile of the entropy rate.}
\label{Fig2}
\end{figure}

To illustrate these points we consider 3-state processes taken randomly in each cumulated quintile of the process entropy rate distribution. Figure \ref{Fig2} shows the effectiveness of our approach by highlighting that processes having large entropy rate are more suited to our approach. In addition, we observe that $\mu(50)$ seems to be below 10\%, which seems to infirm our intuition that MaxEnt should perform better for higher-dimensional state spaces.

As long as stationary processes only are considered, the MaxEnt method is actually of limited interest since nothing precludes the use of arbitrarily long samples. Things are very different when the dynamics itself changes over time, for then a quick estimation of dynamical parameters becomes necessary. The crucial point, which follows immediately from our previous results, is as follows: if the coefficients evolve within $M(\tau)$, where $\tau$ is the typical time scale on which the dynamic parameters of the dynamics change, then MaxEnt provides a quicker estimation of the instantaneous dynamics than sampling does.

The change of perspective should be emphasized: while the maximum entropy approach is stationary in essence, we now apply it on non-stationary processes by approximating them locally in time by (hopefully) effective Markov processes.

\begin{figure}
\includegraphics[scale=0.4]{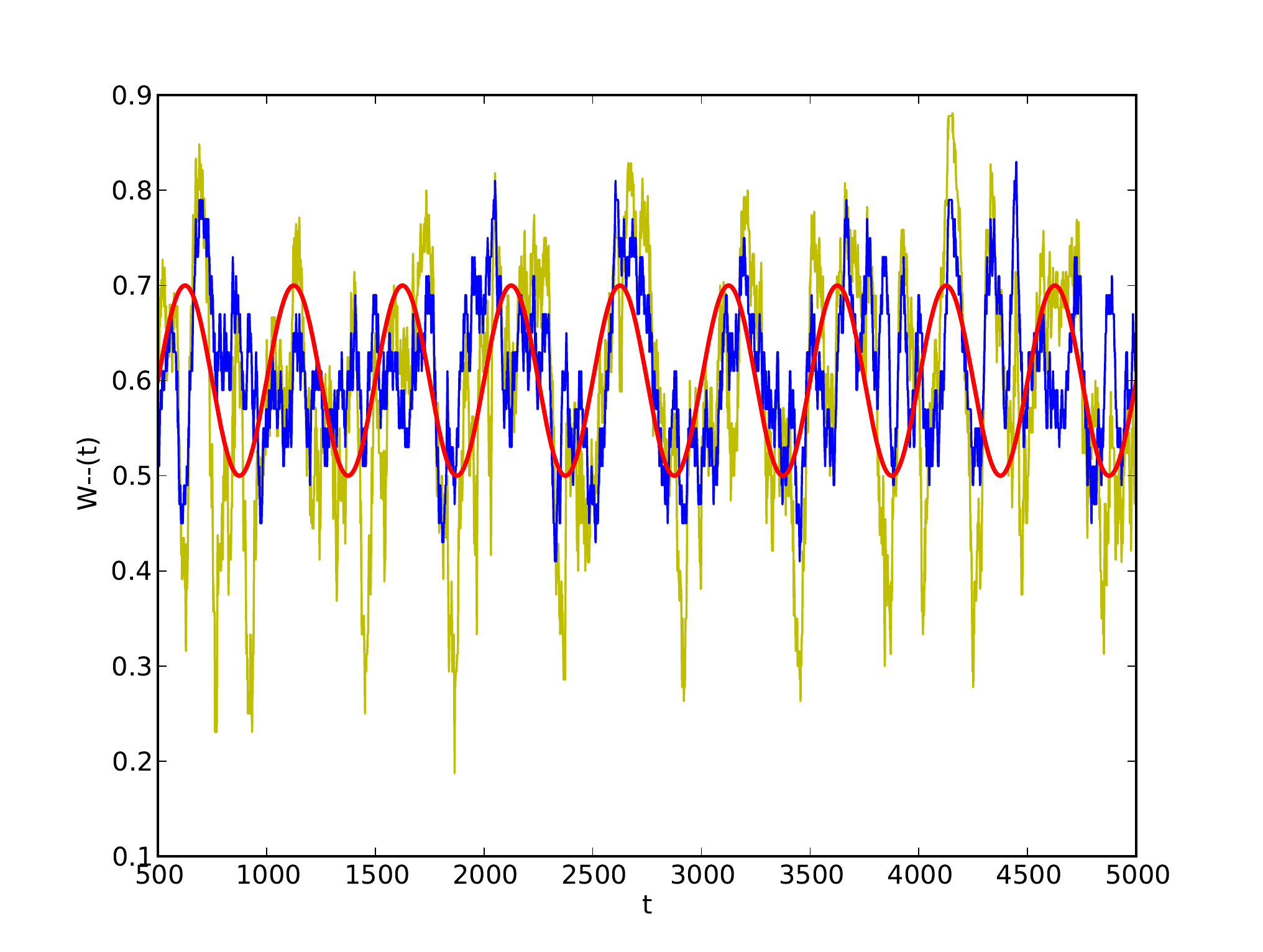}
\caption{\small A comparison of the true coefficient $W_{--}(t)$ (red) with its MaxEnt (blue) and sampling (yellow) estimates, for a realization of the toy process described by \eqref{Toy}.}
\label{Fig3}
\end{figure}

Figure \ref{Fig3} conveys a qualitative illustration of this approach for a 2-state process generated by the time-varying transition matrix

\begin{equation}
\mathbf{W} (t) = \left(\begin{array}{cc}
0.6 + 0.1 \sin \left( \frac{2 \pi t}{T} \right) & 0.4 - 0.1 \sin \left( \frac{2 \pi t}{T} \right) \\
0.4 - 0.1 \sin \left( \frac{2 \pi t}{1.2T} \right) & 0.6 + 0.1 \sin \left( \frac{2 \pi t}{1.2T} \right) \\
\end{array}
\right) ,
\label{Toy}
\end{equation}
where $T=500$. Due to the relatively short period of oscillation,  considering samples more than a few dozens of time units long would give meaningless over-averaged results. The figure shows that for a sliding window of size $n=50$, the MaxEnt estimate (shown in blue) provides a better match of the coefficient $W_{--}(t)$ (red). In particular, it avoids the large deviations shown by the sampling estimate (yellow).

We now turn our attention to an empirical application aiming at quantifying risk of a financial asset and thus at estimating its return tail distributions. For convenience, we choose to consider the EUR/USD price series over the period Jan 1 2009 up to Jan 1 2012. A series of returns is established by defining $r_t=(p_t - p_{t-1})/p_{t-1}$ where $p_t$ is the price at time $t$ and where the data is picked up every 15 minutes. A 3-state time series $x_t$ is built by discretizing returns according to
\begin{equation}
x_t=
\left\{ 
  \begin{aligned}
  -1 & \quad \text{if} \; r_t < -0.01 \% \\ 
  +1 & \quad \text{if} \; r_t > 0.01 \% \\ 
  0  & \quad \text{otherwise}. \\ 
  \end{aligned}
\right.
\label{eq:disc}
\end{equation} 

Modelling the discretized time series with a 3-states Markov chain, we estimate the transition matrix using the MaxEnt method. This allows to compute the distribution $s$ steps ahead of current time $t$ as $q_{t+s}(k \vert i_0)=\sum_{i_1,\cdots,i_s} W_{i_0,i_1} \cdots W_{i_{s-1},i_s}$ with $k=i_1+\cdots+i_s$ and where $i_\tau \in \{-1,0,+1\}$ is the state at time $t+\tau$.

\begin{figure}
\includegraphics[scale=0.5]{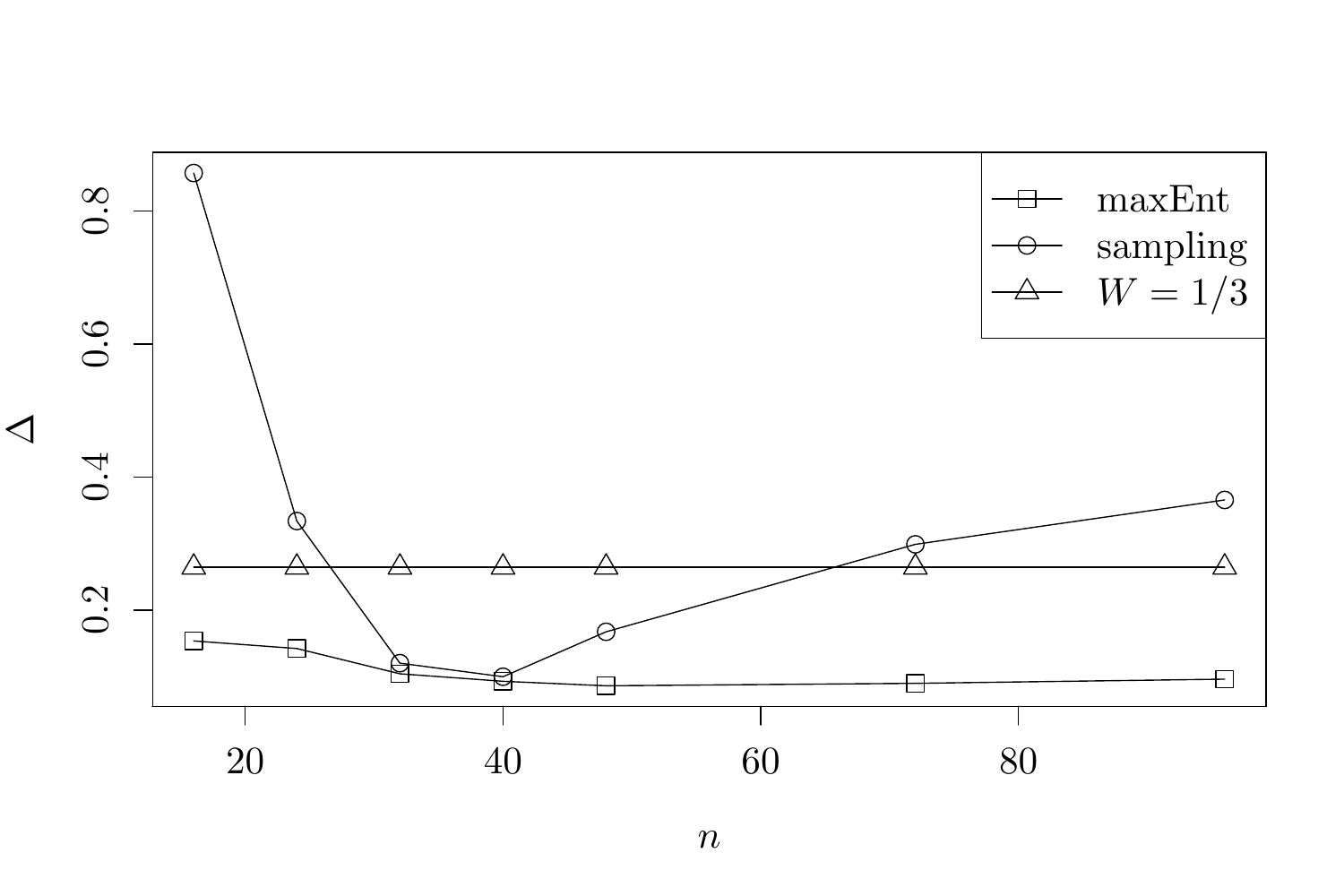}
\caption{\small The average error of fat tail distribution estimations, as a function of the sample size. MaxEnt (squares) and sampling (circles) approaches are used, as well as the naive guess assuming all equiprobable transitions, $W_{ij}=1/3$ $\forall i,j$ (triangles).}
\label{Fig4}
\end{figure}

In order to assess the quality of the tails of the distribution $q$, we define its first and last ten symmetrized \footnote{That is we aggregate the first and the last centile, and so on.} centiles $\pi_k$, $k=1, ..., 10$, and compare them to the fraction $\hat{\pi}_k$ of realizations falling actually into the centile. We then compute the average error $\Delta =\sum_{k=1}^{10} |\pi_k - \hat{\pi}_k|/\pi_k$. Figure~\ref{Fig4} shows, as a function of the sample length $n$, the average error $\Delta$ obtained using the MaxEnt (squares) against the sampling (circles) approach. According to our expectations, we observe that MaxEnt outperforms sampling for sample length shorter than $n=40$. Interestingly, we note that for longer samples, sampling errors seem to grow and depart from MaxEnt errors which we believe to be an over-average effect as discussed above. Figure~\ref{Fig4} also shows that a non-informed guess (triangles), assuming all transitions equiprobable, is alway outperformed by MaxEnt. This highlights that a financial process, exhibiting low correlations, appears to be suited to our procedure.

It is remarkable that the MaxEnt method turns out to be relevant in this context, since contrarily to our previous toy dynamics, it is probable that the underlying dynamics does not change smoothly nor evolve in a subspace of the space of stochastic matrices which is suitable to the MaxEnt procedure.

To summarize, we have shown how an adaptation of the maximum entropy method may be applied to Markov processes. This method gives more accurate estimates of the transition parameters when short samples only are available for inference. This is however true only when processes to estimate fit to the structure imposed by the MaxEnt procedure, but we argue that many processes of interest do satisfy this property. The full strength of the method appears when dealing with processes changing over time; then, under the same conditions as in the stationary case, MaxEnt provides a quicker on-the-fly estimation of the dynamical parameters. We illustrated the relevance of this approach on empirical data, for which we got a considerably better estimate of the tails of the forecasted distribution of trajectories compared to results obtained using naive sampling methods. It should be emphasized that our approach is susceptible of considerable improvement by implementing constraints beyond the simple one-step autocorrelation considered here, as long as one is able to deal with the extra complexity of the MaxEnt algorithm.

The authors would like to acknowledge funding from the European Union Seventh Framework Programme, under grant agreement 317534.


\bibliography{References}

\end{document}